\documentclass[12pt]{article}
\usepackage{natbib}
\usepackage{tikz}
\usetikzlibrary{shapes.geometric,fit}
\usepackage{graphicx}      
\usepackage{subcaption}    
\usepackage{amsmath}
\newtheorem{example}{Example} 
\usepackage{soul} 
\usepackage{makecell} 
\usepackage{url}
\usepackage{multirow}
\usepackage{multicol}
\usepackage{authblk}
\usepackage{booktabs}
\usepackage{caption}
 \usepackage{url}

\usepackage{algorithm}
\floatname{algorithm}{Algorithm}
\usepackage{tabularx}
\usepackage{array}
\newcolumntype{L}{>{\raggedright\arraybackslash}X}
\newcolumntype{C}{>{\centering\arraybackslash}X}
\newcolumntype{R}{>{\raggedleft\arraybackslash}X}

\usepackage{multirow}
\usepackage{multicol}
\usepackage{longtable}

\usepackage{authblk}
\usepackage{eurosym}
\usepackage{booktabs}
\usepackage{enumitem}

\usepackage{blindtext}

\usepackage{lscape}
\usepackage{pdflscape}

\usepackage{geometry}
\usepackage{setspace}
\title{A Joint Sampling Design for the Related Populations\\ of Parents and Children}

\author[1]{Yves Till\'e}
\author[2]{Mar\'ia Guadarrama Sanz\footnote{Corresponding author. Luxembourg Institute of Socio-Economic Research (LISER). Address: 11, Porte des Sciences, L-4366 Esch-sur.Alzette/Belval (Luxembourg). Tel: +352 585855605. maria.guadarramasanz@liser.lu}}
\affil[1]{\small Institut de statistique, Universit\'e de Neuch\^atel. Switzerland}
\affil[2]{\small Luxembourg Institute of Socio-Economic Research, Luxembourg}

\date{}

\begin{document}
\maketitle

\begin{abstract}
We propose a novel joint sampling design for surveys involving two interrelated populations of parents and children in the presence of complex family structures, including shared parental responsibility and blended families. The design combines a graph-theoretic representation of co-parenting relationships with balanced sampling. This framework ensures that, in the first sampling phase, exactly one parent is selected from each co-responsibility pair while maintaining accurate inference for both the parent and child populations. A second sampling phase selects one child per sampled parent using unequal probabilities that produce consistent weighting and unbiased estimation. The proposed framework accommodates overlapping family relationships while balancing auxiliary information for both populations. Results based on the Luxembourg population registers show that the estimated totals are virtually identical to the known population totals for all balancing variables,  and other auxiliary variables, demonstrating the excellent precision and efficiency of the proposed sampling design.
\vspace{0.1 cm}
\end{abstract}

\section{Introduction and Recommendations}\label{sec:intro}

The Luxembourgish Ministry of Education aims to understand parental perceptions of non-formal education among children aged 0--12 years in the country. In Luxembourg, non-formal education refers to organized educational and care services that are provided outside of the formal school system. This includes childcare centers, after-school care centers and registered childminders. The focus is on fostering children's development through everyday learning experiences rather than formal instruction or academic assessment.
 The administrative data, currently available, provide insight into who does (or does not) participate in non-formal education, how frequently, and along what pathways. These data have a major limitation: they provide no information on the perceptions, motivations, or daily constraints that influence parental decisions. 
Therefore, to answer the question posed by an attached Ministry's unit, the National Observatory for Children, Youth and School Quality (OEJQS in French) ``How do families perceive non-formal education, its quality, and its benefits?" administrative data alone are insufficient. In a context where the OEJQS seeks to increase participation in non-formal education by making it free and expanding its availability, it is essential to understand why certain families, particularly those with the lowest incomes, still do not participate. Therefore, collecting information directly from parents of children eligible for non-formal education with a survey is important. The study population and the reference population coincide and are defined as all parents of children aged between 0 and 12 years old, regardless of whether they participate in non-formal education and who were living in Luxembourg in December 2025. The parents will answer questions about themselves, their household composition, and one of the children for whom they are responsible.

The requirements of this survey are:
\begin{itemize}
\item to survey all parents who are solely responsible for all of their children,
\item to survey one of the two parents in each pair sharing responsibility for at least one child.
\item Subsequently, one child is selected at random from among all children for whom the selected parent is the sole responsible parent or shares responsibility so that the parent replies the questions targeted to the selected child.
\end{itemize}

The random selection of parents and then children is complex for blended families, where a parent may share responsibility for children with several other parents. Furthermore, a child may have only one responsible parent, who may in turn share responsibility for one or more children with other parents. Each child may have one or two parental relationships, which complicates the management of family relationships. We account for these situations in order to construct a balanced sampling design as defined in \citet{dev:til:04a}. 

The proposed sampling design allows for accurate extrapolation not only for the parent population but also for the child population. Furthermore, the sampling design must be balanced on certain auxiliary variables, for both parents and children, to ensure the accuracy of the estimators.
The survey therefore targets two related populations simultaneously: the population of parents and the population of children. The sampling design is constructed so that valid and balanced estimators can be produced for both populations.

The remainder of the paper is organized as follows. Section~\ref{sec:descr} describes the parent--child relationships observed in the registers and introduces the concepts of co-responsibility pairs and co-responsibility chains. Section~\ref{analreg} describes these relationships in the registers. Section~\ref{sec:notation} presents the notation and formal framework used throughout the paper. Section~\ref{sec:sampselpar} develops the proposed sampling design for the parent population. Section~\ref{secchildr} describes the selection of children and the estimation procedures. Section~\ref{sec:results} presents the sampling results and balancing properties of the design. In Section~\ref{sec:weight}, we provide some tables that allow to asses the quality of the balanced sample.
Finally, Section~\ref{sec:conclusion} concludes the paper.

\section{Description of the Relationships Between Parents and Children}\label{sec:descr}

The sampling frame is constructed from two administrative registers, precisely, the national register of the physical persons (RNPP, in French) and the register of affiliated individuals in the Luxembourgish Social Security, held by the Inspection of Social Security (IGSS, in French). These registers contain socio-demographic information, employment characteristics, gross earnings, social benefits, and parental relationship records \citep[see][for further details]{IGSS2024}. After data cleaning, we encountered 177\,237 parent-child relationships. The resulting database contains 115\,290 parents and 92\,515 children. Children depend on either a single parent, which is the case for 7\,819 children, or two responsible parents, this happens for 84\,696 children. 

A co-responsibility pair is defined as the relationship between two parents who are co-responsible for at least one child. These pairs may overlap (a parent may belong to multiple co-responsibility pairs). It is, then, possible to split the parent population into three groups, according to the type of co-responsibility pair they belong to:
\begin{itemize}
\item \textit{Group~A.} Parents who do not belong to any co-responsibility pair. That is, those who are solely responsible for all their children. All of these parents will be included in the sample.  
\item \textit{Group~B.} Parents who belong to a co-responsibility pair that does not overlap with any other co-responsibility pair. A parent in \textit{Group~B} may be solely responsible for one or more children while also sharing responsibility for one or more other children with a single other parent. From each pair, one of the two parents will be selected at random. 
\item \textit{Group~C.} All parents belonging to a co-responsibility pair that overlaps with at least one other co-responsibility pair. In each pair, one of the two parents is selected at random. 
A parent in \textit{Group~C} may be solely responsible for one or more children while also sharing responsibility for one or more other children with one or more other parents. 
\end{itemize}

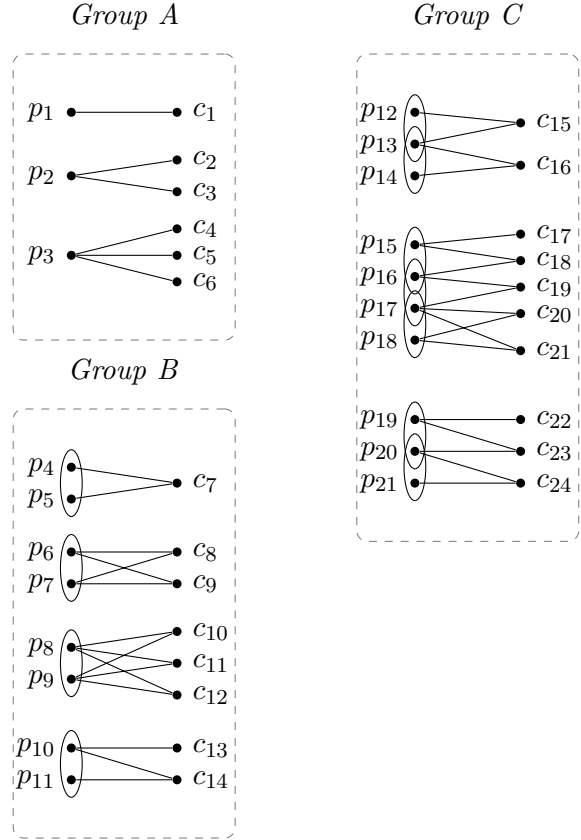
\begin{figure}[htb!]
    \centering
    \begin{tikzpicture}[
        scale=0.70,
        dot/.style={circle, fill=black, inner sep=1.2pt},
        every node/.style={font=\small},
        group/.style={draw, ellipse, inner sep=1pt, thin},
        biggroup/.style={draw, dashed, gray, rounded corners, inner sep=20pt}
    ]
    \begin{scope}
        \node[dot, label=left:$p_1$] (p1) at (0,0) {}; \node[dot, label=right:$c_1$] (e1) at (2,0) {}; \draw (p1) -- (e1);
        \node[dot, label=left:$p_2$] (p2) at (0,-1.2) {}; \node[dot, label=right:$c_2$] (e2) at (2,-0.9) {}; \node[dot, label=right:$c_3$] (e3) at (2,-1.5) {}; \draw (p2) -- (e2) (p2) -- (e3);
        \node[dot, label=left:$p_3$] (p3) at (0,-2.7) {}; \node[dot, label=right:$c_4$] (e4) at (2,-2.2) {}; \node[dot, label=right:$c_5$] (e5) at (2,-2.7) {}; \node[dot, label=right:$c_6$] (e6) at (2,-3.2) {}; \draw (p3) -- (e4) (p3) -- (e5) (p3) -- (e6);
        
        \node[biggroup, fit=(p1) (e6), label={[yshift=5pt]above:\textit{Group~A}}] (boxA) {};

        \begin{scope}[yshift=-2.5cm]
            \node[dot, label=left:$p_4$] (p4) at (0,-4.2) {}; \node[dot, label=left:$p_5$] (p5) at (0,-4.8) {}; \node[dot, label=right:$c_7$] (e7) at (2,-4.5) {}; \draw (p4) -- (e7) (p5) -- (e7); \node[group, fit=(p4) (p5)] {};
            \node[dot, label=left:$p_6$] (p6) at (0,-5.8) {}; \node[dot, label=left:$p_7$] (p7) at (0,-6.4) {}; \node[dot, label=right:$c_8$] (e8) at (2,-5.8) {}; \node[dot, label=right:$c_9$] (e9) at (2,-6.4) {}; \draw (p6) -- (e8) (p6) -- (e9) (p7) -- (e8) (p7) -- (e9); \node[group, fit=(p6) (p7)] {};
            \node[dot, label=left:$p_8$] (p8) at (0,-7.6) {}; \node[dot, label=left:$p_9$] (p9) at (0,-8.2) {}; \node[dot, label=right:$c_{10}$] (e10) at (2,-7.3) {}; \node[dot, label=right:$c_{11}$] (e11) at (2,-7.9) {}; \node[dot, label=right:$c_{12}$] (e12) at (2,-8.5) {}; \draw (p8) -- (e10) (p8) -- (e11) (p8) -- (e12) (p9) -- (e10) (p9) -- (e11) (p9) -- (e12); \node[group, fit=(p8) (p9)] {};
            \node[dot, label=left:$p_{10}$] (p10) at (0,-9.5) {}; \node[dot, label=left:$p_{11}$] (p11) at (0,-10.1) {}; \node[dot, label=right:$c_{13}$] (e13) at (2,-9.5) {}; \node[dot, label=right:$c_{14}$] (e14) at (2,-10.1) {}; \draw (p10) -- (e13) (p10) -- (e14) (p11) -- (e14); \node[group, fit=(p10) (p11)] {};
            
            \node[biggroup, fit=(p4) (e14), label={[yshift=5pt]above:\textit{Group~B}}] (boxB) {};
        \end{scope}
    \end{scope}

    \begin{scope}[xshift=6.5cm]
        \node[dot, label=left:$p_{12}$] (p12) at (0,0) {}; \node[dot, label=left:$p_{13}$] (p13) at (0,-0.6) {}; \node[dot, label=left:$p_{14}$] (p14) at (0,-1.2) {};
        \node[dot, label=right:$c_{15}$] (e15) at (2,-0.2) {}; \node[dot, label=right:$c_{16}$] (e16) at (2,-1) {};
        \draw (p12) -- (e15) (p13) -- (e15) (p13) -- (e16) (p14) -- (e16);
        \node[group, fit=(p12) (p13)] {}; \node[group, fit=(p13) (p14)] {};

        \node[dot, label=left:$p_{15}$] (p15) at (0,-2.5) {}; \node[dot, label=left:$p_{16}$] (p16) at (0,-3.1) {};
        \node[dot, label=left:$p_{17}$] (p17) at (0,-3.7) {}; \node[dot, label=left:$p_{18}$] (p18) at (0,-4.3) {};
        \node[dot, label=right:$c_{17}$] (e17) at (2,-2.3) {}; \node[dot, label=right:$c_{18}$] (e18) at (2,-2.8) {};
        \node[dot, label=right:$c_{19}$] (e19) at (2,-3.3) {}; \node[dot, label=right:$c_{20}$] (e20) at (2,-3.8) {};
        \node[dot, label=right:$c_{21}$] (e21) at (2,-4.5) {};
        \draw (p15) -- (e17) (p15) -- (e18) (p16) -- (e18) (p16) -- (e19) (p17) -- (e19) (p17) -- (e20) (p18) -- (e20) (p17) -- (e21) (p18) -- (e21);
        \node[group, fit=(p15) (p16)] {}; \node[group, fit=(p16) (p17)] {}; \node[group, fit=(p17) (p18)] {}; 

        \node[dot, label=left:$p_{19}$] (p19) at (0,-5.8) {}; \node[dot, label=left:$p_{20}$] (p20) at (0,-6.4) {}; \node[dot, label=left:$p_{21}$] (p21) at (0,-7.0) {};
        \node[dot, label=right:$c_{22}$] (e22) at (2,-5.8) {}; \node[dot, label=right:$c_{23}$] (e23) at (2,-6.4) {}; \node[dot, label=right:$c_{24}$] (e24) at (2,-7.0) {};
        \draw (p19) -- (e22) (p19) -- (e23) (p20) -- (e23) (p20) -- (e24) (p21) -- (e24);
        \node[group, fit=(p19) (p20)] {}; \node[group, fit=(p20) (p21)] {};

        \node[biggroup, fit=(p12) (e24), label={[yshift=5pt]above:\textit{Group~C}}] {};
    \end{scope}

    \end{tikzpicture}
    \caption{Examples of relationships between parents and children according to \textit{Groups A, B}, and~\textit{C}.\label{fig1}}
\end{figure}

Let $U=\{p_1,\ldots,p_k,\ldots,p_N\}$ denote the set of \(N\) parents, and let $V=\{c_1,\ldots,c_i,\ldots,c_M\}$ denote the set of \(M\) children. For each parent \(p_k\), the set of children for whom \(p_k\) is responsible or jointly responsible is denoted by \(P_k\).
Relationships between parents and children can be complex. Figure~\ref{fig1} shows some examples of relationships between parents and children according to the groups in the populations of $N=21$ parents and $M=24$ children. Children $c_{13}$, $c_{17}$, and $c_{22}$ each have a single responsible parent, but their responsible parent is also co-responsible for another child.

 \begin{figure}[htb!]
  \centering
 \includegraphics[width=0.7\textwidth]{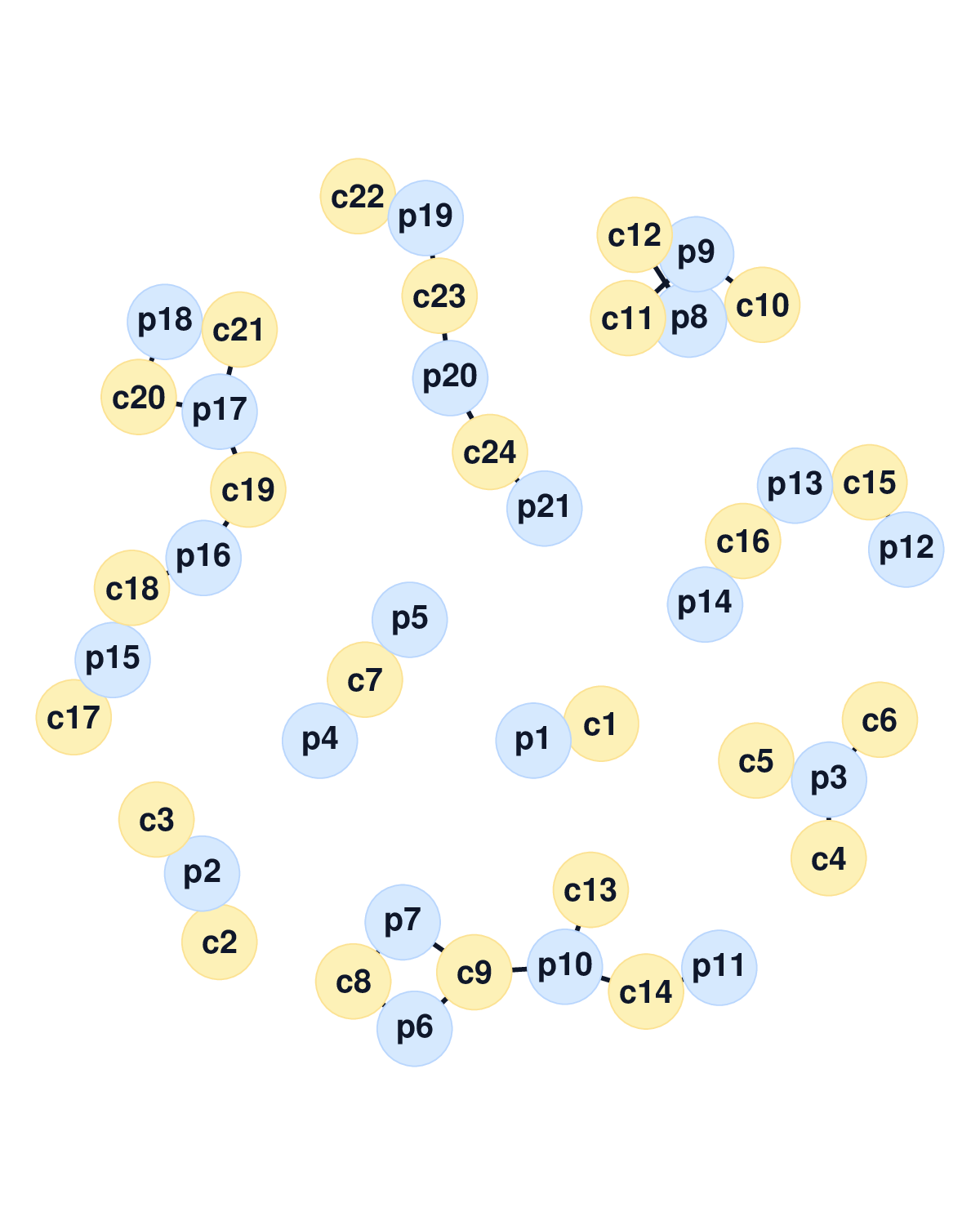}
\vspace{-0.7cm} 
\caption{Graph representation of co-responsibility chains described in Figure \ref{fig1}. Parent nodes ($p$) are shown in blue, while child nodes ($c$) appear highlighted in yellow. \label{Fig2}}
 \end{figure}

Co-responsibility chains are defined as connected components in the sense of graph theory. The co-responsibility chains are thus the largest sets of parents linked to one another by ties of co-responsibility.  Using the example in Figure~\ref{fig1}, we can identify 10 chains of co-responsibility, 3 in \textit{Group~A}  (these are responsibility chains), 4 in \textit{Group~B} and 3 in \textit{Group~C}, which are described in Table~\ref{chaineex}. Each co-responsibility chain consists of a certain number of parents and a certain number of children. It is essential to identify chains of shared responsibility in order to assign parents to groups and for the sample selection described below. 

\begin{table}[htb!]
\centering
\caption{Co-responsibility chain (connected components) related to the example in Figure~\ref{fig1}\label{chaineex}}
\vspace{-0.3cm}
\small
\begin{tabularx}{\textwidth}{LCCCC}
\toprule
Group                     & Number of & Parents & Number of  & Children \\
                          &  Parents          &               & children     & \\
\hline
\textit{Group A}  & 1& $\{p_1\}$ & 1  & $\{c_1\}$ \\
                  & 1& $\{p_2\}$ & 2  & $\{c_2,c_3\}$ \\
                  & 1& $\{p_3\}$ & 3  & $\{c_4,c_5,c_6\}$ \\
\textit{Group B}  & 2& $\{p_4,p_5\}$ & 1  & $\{c_7\}$ \\
                  & 2& $\{p_6,p_7\}$ & 2  & $\{c_8,c_9\}$ \\
                  & 2& $\{p_8,p_9\}$ & 3  & $\{c_{10},c_{11},c_{12}\}$ \\
                  & 2& $\{p_{10},p_{11}\}$ & 2  & $\{c_{13},c_{14}\}$ \\
\textit{Group C}  & 3& $\{p_{12},p_{13},p_{14}\}$ & 2  & $\{c_{15},c_{16}\}$ \\
                  & 4& $\{p_{15},p_{16},p_{17},p_{18}\}$ & 5  & $\{c_{17},c_{18},c_{19},c_{20},c_{21}\}$ \\
                  & 3& $\{p_{19},p_{20},p_{21}\}$ & 3  & $\{ c_{22},c_{23},c_{24} \}$ \\
\bottomrule
\end{tabularx}
\end{table}
\section{Analysis of the Relationship in the Registers}\label{analreg}
We analyzed the RNPP and IGSS registries.  After identifying 177\,237 parent-child relationships, we then constructed all co-responsibility chains. 
The total number of parents (115\,290) is split as follows: There are 5\,386 parents belonging to \textit{Group~A}, 106\,222 parents in \textit{Group~B}, and finally 3\,682 parents in \textit{Group~C}. Parents are grouped into 59\,631  co-responsibility chains. Precisely, there are 5\,386 responsibility chains in \textit{Group~A}, 53\,111 co-responsibility chains in \textit{Group~B}, and 1\,134 co-responsibility chains in \textit{Group~C}. The distribution of parents and children by co-responsibility {chains} is described in Table~\ref{tab:distfam}.

\begin{table}[htb!]
\centering
\caption{Number of co-responsibility chains with respect to the number of children and number of parents\label{tab:distfam}}
\vspace{-0.3cm}
\small
\setlength{\tabcolsep}{6pt}
\begin{tabular}{crrrrrrrr|r|r}
\toprule
\textrm{Number of} & \multicolumn{8}{c|}{\textrm{Number of parents}} & \textrm{Total of}& \textrm{Total of}  \\ \cline{2-9}
\textrm{children}
& \textrm{1} & \textrm{2} & \textrm{3} & \textrm{4}
& \textrm{5} & \textrm{6} & \textrm{7} & \textrm{9}
& \textrm{chains} & \textrm{children} \\
\hline
 \textrm{1}  & 4\,196  & 29\,010 &         &         &       &       &       &       & 33\,206  & 33\,206 \\
 \textrm{2}  & 1\,013  & 19\,660 & 525     &         &       &       &       &       & 21\,198  & 42\,396 \\
 \textrm{3}  & 148     & 3\,742  & 279     & 84      &       &       &       &       & 4\,253   & 12\,759 \\
\textrm{4}  & 25      & 590     & 93      & 62      & 15    &       &       &       & 785     & 3\,140 \\
\textrm{5}  & 4       & 85      & 13      & 21      & 19    & 1     &       &       & 143     & 715 \\
\textrm{6}  &         & 20      & 3       & 4       & 3     & 2     & 1     &       & 33      & 198 \\
\textrm{7}  &         & 4       &         & 1       & 4     &       &       &       & 9       & 63 \\
\textrm{8}  &         &         &         &         &       & 1     &       &       & 1       & 8 \\
\textrm{9}  &         &         &         &         &       &       & 1     &       & 1       & 9 \\
\textrm{10} &         &         & 1       &         &       &       &       &       & 1       & 10\\
\textrm{11} &         &         &         &         &       &       &       & 1     & 1       & 11\\
\hline
\textrm{Total of chains} &
 5\,386 & 53\,111 & 914 & 172 & 41 & 4 & 2 & 1 & 59\,631 & 92\,515 \\
\hline
\textrm{Total of parents } &
5\,386 & 106\,222 & 2\;742 & 688 & 205 & 24 & 14 & 9 & 115\,290 & \\
\hline
& \textit{Group~A} & \textit{Group~B} & \multicolumn{6}{c}{\textit{Group~C}} & \multicolumn{1}{c}{} & \multicolumn{1}{c}{} \\
\cline{1-9}
\end{tabular}
\end{table}

The interpretation of Table~\ref{tab:distfam} is as follows. The column corresponding to one parent belongs to \textit{Group~A}. The column corresponding to two parents belongs to \textit{Group~B}, and the columns corresponding from three to nine parents belong to \textit{Group~C}. Table \ref{tab:distfam} reveals that the most common situation consists of two parents and one child (29\,010), followed by two parents and two children (19\,660), and then, one parent and one child (4\,196).  There are 3\,682 ($2\,742 + 688 + 205 + 24  + 14+ 9$) parents who are part of a co-responsibility chain involving more than two adult people. The largest co-responsibility chain  consists of 9 parents and 11 children. 

\begin{figure}[htb!]
    \centering\hspace{-0.5cm}
    \begin{subfigure}[b]{0.4\textwidth}
        \centering        \includegraphics[width=\textwidth]{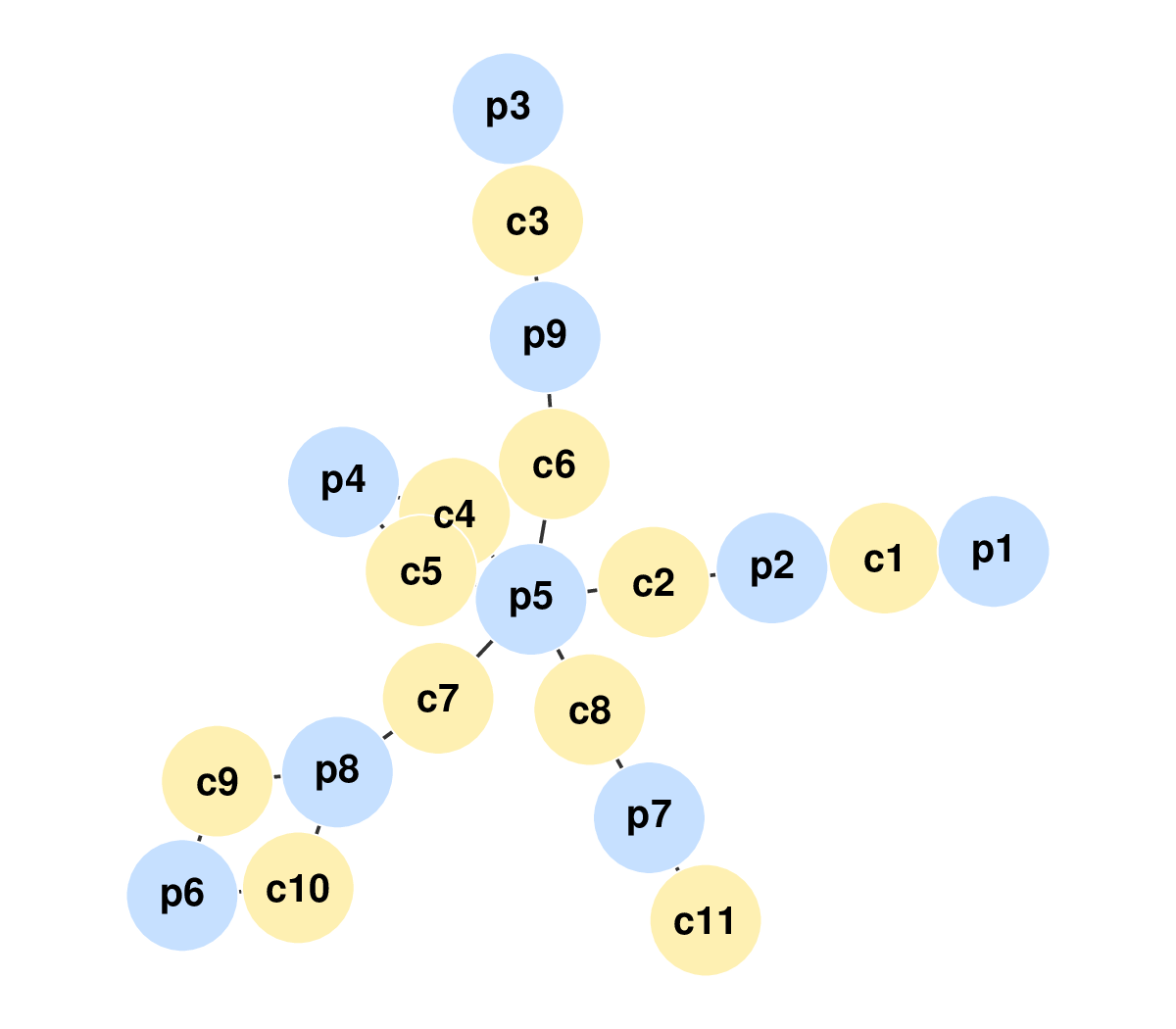}
    \end{subfigure}
    \hspace{-1.8cm}
    \begin{subfigure}[b]{0.4\textwidth}
        \centering
\includegraphics[width=\textwidth]{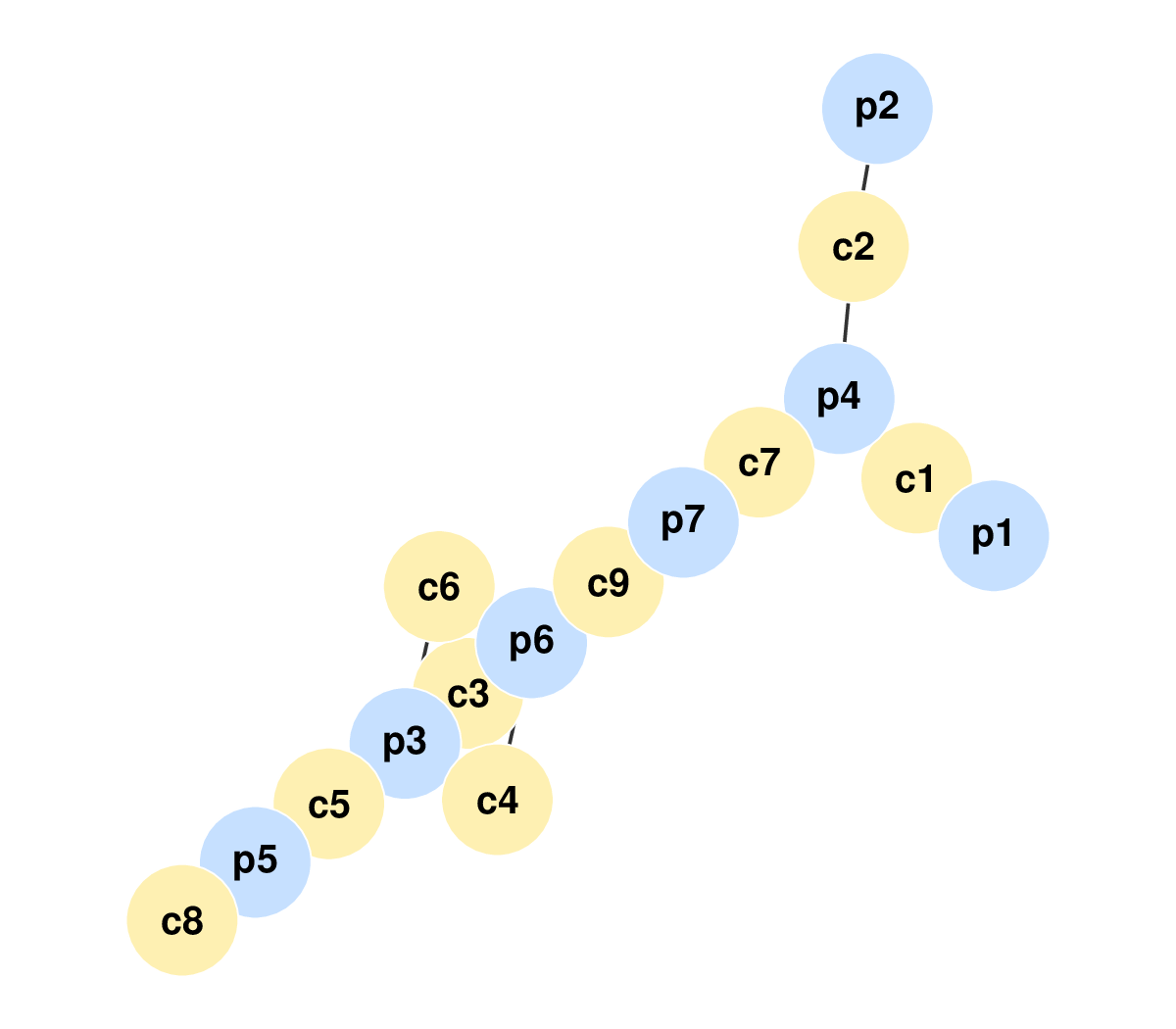}
    \end{subfigure}
    \hspace{-2.3cm}
    \begin{subfigure}[b]{0.4\textwidth}
        \centering     \includegraphics[width=\textwidth]{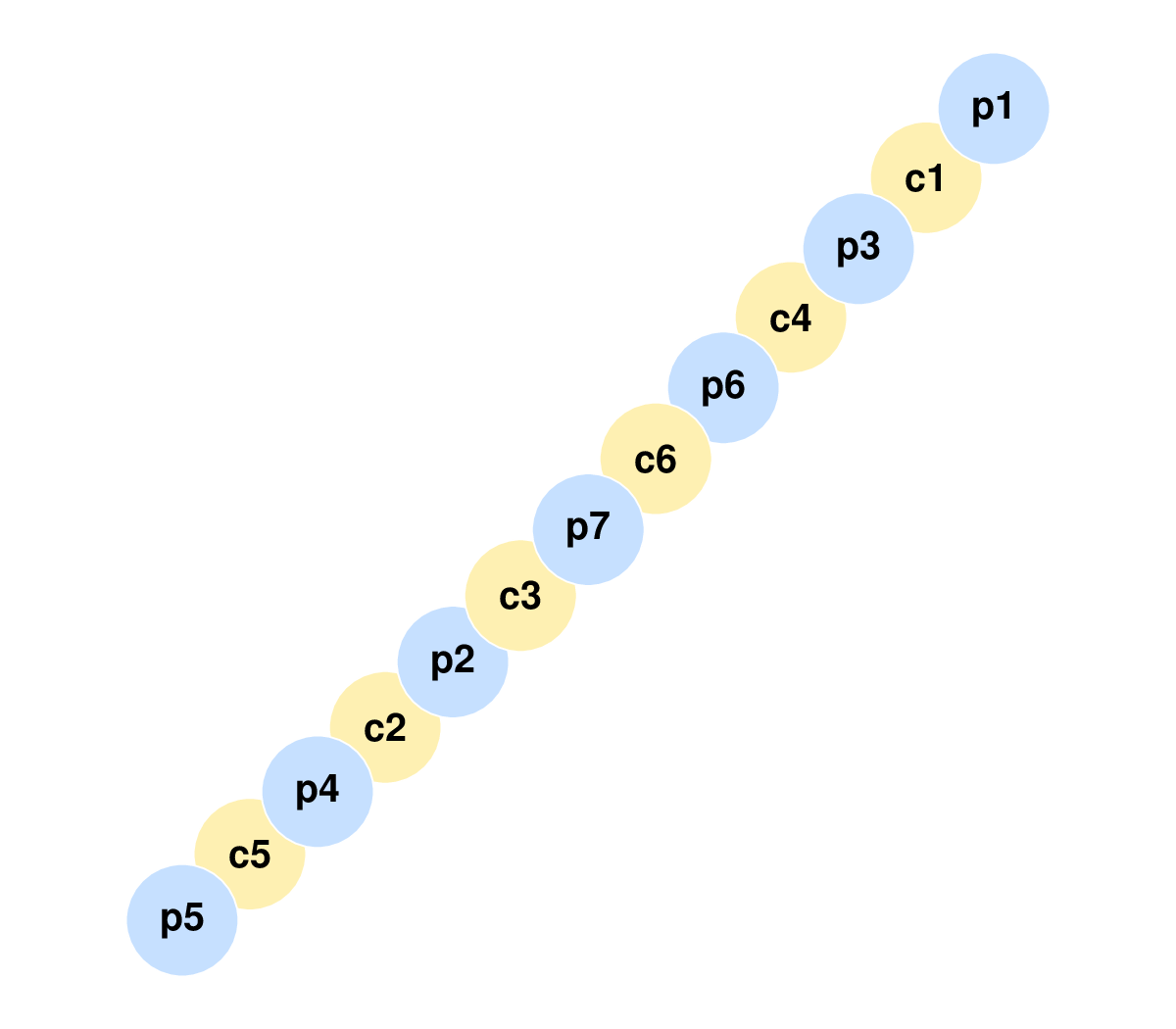}
    \end{subfigure}
    \caption{Co-responsibility chains displaying nine parents (left) and seven parents (middle and right figures) within the sampling frame. Parent nodes ($p$) are shown in blue, while child nodes $c$ appear highlighted in yellow}
    \label{fig:familias_redes}
\end{figure}

Table \ref{tab:distfam} indicates that most family structures are simple co-responsibility pairs, with one or two parents, while complex overlapping co-responsibility chains remain relatively rare being around $6\%$ of the cases. Figure \ref{fig:familias_redes} plots the largest co-responsibility chains with 9 parents, on the left, and 7 parents, on the middle and on the right, respectively. These are the most complicated co-responsibility chains that we have encountered for the survey. In addition to the large amount of parents present in the co-responsibility chain, there are children with only one responsible parent. Precisely, child $c_{11}$ in Figure \ref{fig:familias_redes} left, and child $c_8$ in Figure \ref{fig:familias_redes} center. Because of confidentiality rules, we cannot disclose any further information on the characteristics of the parents nor the children.

\section{Notation and Formalization}\label{sec:notation}

For each parent $k$ in the parent population $U = \{p_1, \dots, p_N\}$, we assume that a vector of auxiliary variables  ${\mathbf{x}}_k$ is available  (such as age group, sex, social security enrollment status, and nationality). For each child $i$ in the child population $V = \{c_1, \dots, c_M\}$, we also assume that we know a vector ${\mathbf{z}}_i$ of auxiliary variables (such as age group, sex, and nationality). The set of children for whom parent $k$ is responsible or jointly responsible is denoted by $P_k$.

Since a child with two responsible parents appears in the records of both parents, its auxiliary information must be split equally between them to avoid double counting.
The aggregated child auxiliary information associated with parent $k$ is defined by
\[
\check{\mathbf{z}}_k = \sum_{i \in P_k} \frac{\mathbf{z}_i}{r_i},
\]
where $r_i$ is the number of parents responsible for child $i$ (1 or 2). 

The child-level auxiliary variables are divided by two whenever the child has two responsible parents. This allows us to preserve the following equality:
\[
\sum_{k \in U} \check{\mathbf{z}}_k = \sum_{k \in U} \sum_{i \in P_k} \frac{\mathbf{z}_i}{r_i} = \sum_{i \in V} \mathbf{z}_i.
\]
This transformation enables child-level auxiliary information to be incorporated into the parent-level balancing equations. By summing the $\check{\mathbf{z}}_k$ over~$U$, we obtain exactly the sum of the $\mathbf{z}_i$ over~$V$.

\section{Sampling Design for Parents}\label{sec:sampselpar}

In \textit{Group A}, we select all the parents. In \textit{Groups B} and \textit{C}, we select a sample of parents, balanced across auxiliary variables, such that only one parent is selected from each co-responsibility pair. This is a problem of overlapping stratification, in which pairs form the strata and one parent is selected from each pair.

In the first step, the inclusion probabilities $\pi_k$ for each parent are therefore 1/2 or 1. We select a balanced sample or approximately balanced sample of parents that satisfies the balancing equations
\begin{equation}
\sum_{k \in U} \frac{\mathbf{x}_k}{\pi_k} I_k \approx \sum_{k \in U} \mathbf{x}_k
\mbox{ and } \sum_{k \in U} \frac{\check{\mathbf{z}}_k}{\pi_k} I_k \approx \sum_{k \in U} \check{\mathbf{z}}_k = \sum_{i \in V} \mathbf{z}_i,
\label{eq1}
\end{equation}
where $I_k$ is 1 if parent $k$ is selected and 0 otherwise. Furthermore, if parents $k$ and $\ell$ are in the same pair, then $I_k + I_\ell = 1$. Co-responsibility pairs are therefore strata in which exactly one parent is always selected. The main difficulty is that co-responsibility pairs may overlap.

In practice, sampling is carried out as follows:
\begin{itemize}
\item
In \textit{Group~A}, all parents are selected with inclusion probability of 1.
\item
In \textit{Group~B}, a balanced and highly stratified sample \citep{hasl:till:2014,eus:jaus:til:2021} is selected with inclusion probabilities equal to 1/2. The strata correspond to pairs of co-responsibility. We include the balancing variables for parents $\mathbf{x}_k$ and children $\check{\mathbf{z}}_k$.
\item
In \textit{Group~C}, we identify all co-responsibility chains (connected components in the graph-theoretic sense) which are the largest sets of parents linked together by ties of co-responsibility. Connected components may contain cycles. A connected component admits a partition into two clusters such that exactly one parent is selected from each co-responsibility pair if and only if the component is bipartite. We thus assume that every connected component is bipartite, equivalently that it contains no odd cycle. In this case, each connected graph can be split into two subsets, which form two clusters. By traversing each connected component, we alternately assign parents to one cluster or the other.
If, within a connected component, all pairs consist of parents of different genders, this method simply involves placing all women in one cluster and all men in the other.
We then implement a balanced stratified cluster sampling design with inclusion probabilities of 1/2~: each stratum corresponds to a connected component and contains the two associated clusters. Balancing can be achieved by summing up the values of the auxiliary variables of the parents $\mathbf{x}_k$ and the children $\check{\mathbf{z}}_k$ within each cluster.
\end{itemize}

\begin{example}
 From Figure~\ref{fig1}, and Table ~\ref{chaineex}, in \textit{Group~C}, there are three connected components: $\{p_{12},p_{13},p_{14}\}$,
$\{p_{15},p_{16},p_{17}, p_{18}\}$, and $\{p_{19},p_{20},p_{21}\}$. We partition each connected component into two groups:
\begin{itemize}
\item
$\{\{p_{12},p_{14}\},\{p_{13}\}\}$,
\item $\{\{p_{15},p_{17}\},\{p_{16}, p_{18}\}\}$ and 
\item $\{\{p_{19},p_{21}\},\{p_{20}\}\}$.
\end{itemize}
In each connected component, one of the two groups is selected with probability 1/2.
\end{example}

\begin{example}
If the connected component contains a cycle of four pairs $\{\{p_1,p_2\},$ $\{p_2,p_3\},$ $\{p_3,p_4\},$ $\{p_4,p_1\}\}$, the two groups are $\{\{p_1,p_3\},\{p_2,p_4\}\}$. By selecting one of the two groups, we select only one parent per pair.
\end{example}

In \textit{Group~C}, a problem could arise when there is a circular chain of pairs containing an odd number of parents, which necessarily implies that at least one pair consists of two parents of the same gender. 
\begin{example}
Suppose we have three parents $\{p_1, p_2, p_3\}$ belonging to the three co-responsibility pairs $\{p_1, p_2\}$, $\{p_2, p_3\}$, and $\{p_3, p_1\}$. This results in an odd cycle of length 3. It is impossible to select exactly one parent from each pair. Indeed, if we select $p_1$, we cannot select $p_2$, which forces us to select $p_3$; we then select both parents from the pair $\{p_3, p_1\}$. Conversely, if we do not select $p_1$, we must select $p_2$, which prevents us from selecting $p_3$; in this case, no parent is selected from the pair $\{p_3, p_1\}$.
This shows that it is impossible to simultaneously satisfy the “one parent per pair” constraint in a connected component containing an odd cycle. \end{example}
In the present application, such a situation would imply the existence of at least one same-gender parental pair. No odd cycle was observed among the 59\,631 connected components identified in the registers.

We proceeded as follows: 
\begin{itemize}
\item Parents in \textit{Group~B} and \textit{Group~C} are partitioned into two clusters within each co-responsibility chain. 

More specifically, in \textit{Group~B}, the clusters have a size 1. We assigned the man in cluster 1 and the woman in cluster 2. For the cases in which there are two men (35 pairs) or two women (168 pairs), we assigned (indistinct) one of the parents to cluster 1 and the other to cluster 2. 

In \textit{Group~C}, we proceeded as follows. When, in a chain of co-responsibility, all pairs consist of people of different genders, all men are assigned to one cluster and all women to the other cluster.
In \textit{Group~C}, we encountered two co-responsibility chains in which a pair of co-responsible individuals consists of two parents of the same gender. In both cases, the clusters were created manually to ensure that the two co-responsible parents of a child do not appear in the same cluster.


The distribution of the number of parents by \textit{Group} and cluster is described in Table~\ref{tab:groups_clusters}. Note that for \textit{Group~A} assign the parents to cluster 1 or cluster 2 is indifferent as we select them all.

\begin{table}[htb!]
    \centering
    \caption{Distribution of parents by \textit{Group} and cluster}
    \vspace{-0.3cm}
    \label{tab:groups_clusters}
    \begin{tabular}{lrr}
 \toprule
     & \multicolumn{2}{c}{\textit{Cluster}} \\
    \cmidrule{2-3}
    \textit{Group} & \textit{1} & \textit{2} \\
    \hline
    \textit{A} & 5\,386  &      \\
    \textit{B} & 53\,111 & 53\,111 \\
    \textit{C} & 1\,763  & 1\,919  \\
    \bottomrule
    \end{tabular}
\end{table}
    
\item    We have defined the following balanced variables.
    \begin{itemize}
        \item[1.] At the level of the parents, we consider the following variables $\mathbf{x}_k$ (1) nationality being Luxembourger or not, and (2) the affiliation status to Luxembourgish social security. Precisely, a) being in full-time parental leave or blue collar; b) white collar; c) civil servant; d) self-employed, unemployed (receiving unemployment benefits), pre-retired, invalid, retired or widow (receiving a surviving spouse pension); e) voluntary insured (paying one's own social security contributions, but as a category other than that of a self-employed worker), not affiliated into social security (that happens when a person is working for (or retired from) an international or foreign institution, or is in a mission in Luxembourg for more than three months); f) co-insured (insured in the name of someone else). 
         \item[2.] At the level of the children, we cross the following variables,  (1) gender; (2) three age class, a) 0-4 years old; b) 5-7 years old; and c) 8-12 years old. Therefore, we have 6 auxiliary variables, $\mathbf{z}_i$.
    \end{itemize}
\end{itemize}
These variables were selected because they are available in the administrative registers and are expected to be associated with both survey participation and the use of non-formal education services.
Table~\ref{tab:population_distr} details the population totals of parents and children with respect to the auxiliary variables used for the sample draw, for the whole population, and within each \textit{Group}.

\begin{table}[ht!]
\small
\centering
\caption{Population distribution of parents and children by nationality (for parents) social security affiliation status (for parents), and by gender and age group (for children), for the whole population and within each \textit{Group}.}\label{tab:population_distr}
\vspace{-0.3cm}
\begin{tabularx}{\textwidth}{llcRRRR}
\toprule
 & \multicolumn{2}{c}{\textrm{Auxiliary variable}} & \textrm{All} & \textrm{\textit{Group A}} & \textrm{\textit{Group B}} & \textrm{\textit{Group C}} \\
\midrule
&  Parents total & & 115\,290 & 5\,386 & 106\,222 & 3\,682\\
\cmidrule{2-7}
& \multirow{2}{*}{\textrm{Nationality}} & Luxembourger & 49\,832 & 1\,932 & 45\,853 & 2\,047 \\
 &  & Not Luxembourger & 65\,458 & 3\,454 & 60\,369 &  1\,635\\
\cmidrule{2-7}
 & \multirow{8}{*}{\makecell[l]{\textrm{Social security} \\ \textrm{affiliation status}}} & Parent, blue collar & 23\,129 & 1\,278 & 20\,833 & 1\,018\\
 &  & White collar & 46\,799 & 1\,850 & 43\,619 & 1\,330 \\
 &  & Civil servant & 13\,028 & 228 &  12\,373 & 427 \\
 &  & \makecell[c]{Self-employed, unemployed, \\ pre-retired, invalid, retired, widow} & 11\,628 & 555 & 10\,692 & 381 \\
 &  & Voluntary insured, Non-affiliated & 11\,841 & 1\,277 & 10\,161 & 403 \\
 &  & Co-insured & 8\,865 & 198 &  8\,544 &  123\\
\midrule
&  Children total & & 92\,515 & 6\,786 & 82\,489 & 3\,240\\
\cmidrule{2-7}
& \multirow{3}{*}{\textrm{Boy}} & 0-4 years old & 17\,365 & 1\,048 & 15\,644 & 673\\
 &  & 5-7 years old & 10\,976 & 775 & 9\,850 &  351\\
 &  & 8-12 years old & 18\,907 & 1\,683 & 16\,517 & 707 \\
\cmidrule{2-7}
 & \multirow{3}{*}{\textrm{Girl}} & 0-4 years old & 16\,511 & 965 & 14\,931 & 615\\
 &  & 5-7 years old & 10\,607  & 716 & 9\,581  & 310  \\
 &  & 8-12 years old & 18\,149 & 1\,599 & 15\,966 &584 \\
\bottomrule
\end{tabularx}
\end{table}

We have selected a total of 60\,338 parents, precisely 5\,386 parents from \textit{Group A}, 53\,111 parents from \textit{Group B} and 1\,841 parents from \textit{Group C}.  

\section{Selection of Children}\label{secchildr}

In the second phase, a single child of each selected parent is selected at random. 
The procedure is as follows:
\begin{itemize}
\item For parents in \textit{Group~A}, one child is selected with equal probability. In this case, let $m_k$ denote the number of children of parent $k$. The probability of selecting a child is therefore $q_i = 1/m_k$, where $i \in P_k$. The weights of the children in Group A are thus
$$
w_i = \frac{1}{\pi_k q_i} = m_k, i\in P_k, k \in \mbox{\textit{Group~A}}.
$$
\item For parents in \textit{Groups~B} and \textit{Group~C}, we select a single child for each selected parent with unequal probabilities.
Children with a single responsible parent whose parent also shares responsibility for another child are assigned twice the selection probability of children with two responsible parents. There are 451 children with a single responsible parent in \textit{Group~B} and 67 children in \textit{Group~C}, which is $0.55~\%$ of the cases in \textit{Group~B} and  $2.11~\%$ of the cases in \textit{Group~C} respectively.
Let
$$
m_k = 2 \sum_{i \in P_k} \frac{1}{r_i}
$$
be the number of children of parent $k$, counting twice those children who have only one responsible parent and whose responsible parent is also co-responsible for at least one other child.
Next, we calculate the unequal probabilities
$$
q_i = \frac{2}{r_i m_k}, i\in P_k.
$$
Thus, we have $\sum_{i \in P_k} q_i = 1.$
\end{itemize}

From Expression~\eqref{eq1}, we have the balancing equations of the first phase:
$$
 \sum_{k \in U} \frac{I_k}{\pi_k}  \sum_{i \in P_k} \frac{\mathbf{z}_i}{r_i} = \sum_{k \in U} \frac{\check{\mathbf{z}}_k}{\pi_k} I_k \approx  \sum_{k \in U} \check{\mathbf{z}}_k = \sum_{i \in V} \mathbf{z}_i,
$$

We again select a highly stratified and balanced design such that
\begin{equation}
\sum_{k \in U} \frac{I_k}{\pi_k} \sum_{i \in P_k} \frac{\mathbf{z}_i J_i}{q_i r_i}
\approx \sum_{k \in U} \frac{I_k}{\pi_k} \sum_{i \in P_k} \frac{\mathbf{z}_i}{r_i},
\label{eq3}
\end{equation}
where $J_i$ is 1 if child $i$ is selected and 0 otherwise. 
We must therefore select a sample of children with inclusion probabilities $q_i$ by balancing on the auxiliary variable $\mathbf{z}_i/(\pi_k r_i)$.
Since
\[
\sum_{i \in P_k} J_i = 1, \mbox{ for all } k \in U,
\]
we must also stratify by the $P_k$. 
Balancing will ensure that the estimates of the totals for the children's auxiliary variables are equal to the population totals.

The weights of the children are as follows:
\begin{itemize}
\item In \textit{Group A} 
$$
\frac{1}{\pi_k q_i r_i} = \frac{m_k}{1 \cdot  1} = m_k, i\in P_k, k \in \mbox{\textit{Group~A}}.
$$
\item
In \textit{Groups B} and \textit{C}:
$$
w_i = \frac{1}{\pi_k q_i r_i} = 2 \frac{r_i \; m_k}{2 r_i }= m_k, i\in P_k, k \in \mbox{\textit{Group~B or Group~C}}.
$$
\end{itemize}

Equation~\eqref{eq3}, can also be written
$$
\sum_{k \in U} I_k m_k  \sum_{i \in P_k} \mathbf{z}_i J_i
= \sum_{k \in U} \frac{I_k}{\pi_k} \sum_{i \in P_k} \frac{\mathbf{z}_i}{r_i}.
$$

This is a two-phase design, in which the second phase depends on the first. The weights of the children are therefore always
\[
w_{i} = m_k, i\in P_k, k\in U.
\]
These weights are random, since they vary depending on which parent the child was selected from. We can estimate $Y$ using
$$
\widehat{Y} = \sum_{i\in V} w_i y_i J_i.
$$

\begin{example}
Consider a selected parent $k$ who is responsible for two children, $i_1$ and $i_2$. Child $i_1$ has only one responsible parent, whereas child $i_2$ has two responsible parents. Thus $r_{i_1}=1$ and $r_{i_2}=2$.

\[
m_k
=
2\left(\frac{1}{r_{i_1}}+\frac{1}{r_{i_2}}\right)
=
2\left(1+\frac{1}{2}\right)
=
3.
\]

The child-selection probabilities are
\[
q_{i_1}
=
\frac{2}{m_k r_{i_1}}
=
\frac{2}{3},
\qquad
q_{i_2}
=
\frac{2}{m_k r_{i_2}}
=
\frac{1}{3}.
\]

The parent $k$ must belong to either \textit{Group B} or \textit{Group C}, so $\pi_k=1/2$. The resulting weights of the children are

\[
w_{i_1}
=
\frac{1}{(1/2)(2/3)(1)}
=
3,
\]

and

\[
w_{i_2}
=
\frac{1}{(1/2)(1/3)(2)}
=
3.
\]

Therefore, both selected children receive the same weight $m_k=3$, illustrating the general result
\[
w_i=\frac{1}{\pi_k q_i r_i}=m_k.
\]
\end{example}

\section{Sampling Results and Balancing Quality}\label{sec:results}

After selecting the sample, we estimated the totals of the auxiliary variables and compared them with the corresponding population totals. 
Table~\ref{tab:population_distr} presents the estimated totals obtained after the first sampling phase. More precisely, we computed
\[
\sum_{k \in U} \frac{\mathbf{x}_k}{\pi_k} I_k
\quad \text{and} \quad
\sum_{k \in U} \frac{\check{\mathbf{z}}_k}{\pi_k} I_k,
\]
as defined in Equation~\eqref{eq1}. These estimates can be compared with the true population totals reported in Table~\ref{tab:groups_clusters}. The estimated totals are almost identical to the corresponding population totals, indicating that the first-phase sample reproduces the balancing variables remarkably well.

Table~\ref{tab:HTestimation} reports the estimated totals for the variables $\mathbf{z}_i$ after the second sampling phase. Specifically, we computed
\[
\sum_{k \in U} \frac{I_k}{\pi_k}
\sum_{i \in P_k}
\frac{\mathbf{z}_i J_i}{q_i r_i},
\]
as defined in Equation~\eqref{eq3}. The results can be compared both with the true population totals in Table~\ref{tab:groups_clusters} and with the intermediate estimates reported in Table~\ref{tab:population_distr}. Once again, the estimated totals are almost identical to the corresponding population totals.

Table~\ref{tab:population_distr} presents the population totals. Tables~\ref{tab:HTestimation} and~\ref{tab:HTchildselect} report the estimated totals after the first sampling stage and the second selection stage, respectively. These results clearly demonstrate the excellent balancing properties of the proposed sampling design. The selected samples reproduce the balancing variables with almost perfect accuracy. As a result, the estimators computed from the samples achieve a high level of precision.

\begin{table}[ht!]
\centering
\caption{Estimation of the distribution of parents and children by nationality (for parents) social security affiliation status (for parents), and by gender and age group (for children), for the whole population and within each \textit{Group}.}\label{tab:HTestimation}
\vspace{-0.3cm}
\begin{tabularx}{\textwidth}{llcRRRR}
\toprule
 & \multicolumn{2}{c}{\textrm{Auxiliary variable}} & \textrm{All} & \textrm{\textit{Group A}} & \textrm{\textit{Group B}} & \textrm{\textit{Group C}} \\
\midrule
& Parents total & & 115\,290 & 5\,386 & 106\,222 & 3\,682\\
\cmidrule{2-7}
& \multirow{2}{*}{\textrm{Nationality}} & Luxembourger & 49\,832 & 1\,932 & 45\,854 & 2\,046 \\
 &  & Not Luxembourger & 65\,458 & 3\,454 & 60\,368 &  1\,636\\
\cmidrule{2-7}
 & \multirow{8}{*}{\makecell[l]{\textrm{Social security} \\ \textrm{affiliation status}}} & Parent, blue collar & 23\,128 & 1\,278 & 20\,832 & 1\,018\\
 &  & White collar & 46\,798 & 1\,850 & 43\,620 & 1\,328 \\
 &  & Civil servant & 13\,030 & 228 &  12\,374 & 428 \\
 &  & \makecell[c]{Self-employed, unemployed, \\ pre-retired, invalid, retired, widow} & 11\,629 & 555 & 10\,692 & 382 \\
 &  & Voluntary insured, Non-affiliated & 11\,841 & 1\,277 & 10\,160 & 404 \\
 &  & Co-insured & 8\,864 & 198 &  8\,544 &  122\\
\midrule
& Children total & & 92\,518 & 6\,786 & 82\,488 & 3\,244\\
\cmidrule{2-7}
& \multirow{3}{*}{\textrm{Boy}} & 0-4 years old & 17\,364 & 1\,048 & 15\,644 & 672\\
 &  & 5-7 years old & 10\,976 & 775 & 9\,850 &  351\\
 &  & 8-12 years old & 18\,906 & 1\,683 & 16\,517 & 706 \\
\cmidrule{2-7}
 & \multirow{3}{*}{\textrm{Girl}} & 0-4 years old & 16\,511 & 965 & 14\,931 & 615\\
 &  & 5-7 years old & 10\,608  & 716 & 9\,581  & 311  \\
 &  & 8-12 years old & 18\,153 & 1\,599 & 15\,965 &589 \\
\bottomrule
\end{tabularx}
\end{table}

\begin{table}[ht!]
\centering
\caption{Estimation of the distribution of children gender and age, for the whole population and within each \textit{Group}.}\label{tab:HTchildselect}
\vspace{-0.3cm}
\begin{tabularx}{\textwidth}{llcRRRR}
\toprule
 & \multicolumn{2}{c}{\textrm{Auxiliary variable}} & \textrm{All} & \textrm{\textit{Group A}} & \textrm{\textit{Group B}} & \textrm{\textit{Group C}} \\
\midrule
&  Children total & & 92\,518 & 6\,786 & 82\,488 & 3\,244\\
\cmidrule{2-7}
& \multirow{3}{*}{\textrm{Boy}} & 0-4 years old & 17\,367 & 1\,050 & 15\,645 & 672\\
 &  & 5-7 years old & 10\,978 & 776 & 9\,851 &  351\\
 &  & 8-12 years old & 18\,907 & 1\,684 & 16\,516 & 707 \\
\cmidrule{2-7}
 & \multirow{3}{*}{\textrm{Girl}} & 0-4 years old & 16\,511 & 963 & 14\,931 & 617\\
 &  & 5-7 years old & 10\,605  & 715 & 9\,579  & 311  \\
 &  & 8-12 years old & 18\,150 & 1\,598 & 15\,966 &586 \\
\bottomrule
\end{tabularx}
\end{table}

We then examined the distribution of parents and children across the different cantons of Luxembourg. Although canton was not included as a balancing variable during the sample selection process, the resulting sample closely reflects the geographical distribution of the target population. Table~\ref{tab:HTcanton} presents the distribution of parents and children by canton for both the general population and the final selected sample. As shown, the sample distributions closely match those of the population, indicating that geographic distribution was preserved despite canton not being explicitly considered in the balancing procedure.

\begin{table}[!ht]
\centering
\caption{Estimation of the distribution of parent and children  by residence canton for the whole population and within each \textit{Group}.}
\label{tab:HTcanton}
\vspace{-0.3cm}
\resizebox{\textwidth}{!}{    \begin{tabular}{lrrrrrrrr}
\toprule
        ~ & \multicolumn{4}{c}{Population} &  \multicolumn{4}{c}{Sample} \\
     \cmidrule{2-9} 
      ~ & All &\textit{Group~A} & \textit{Group~B }& \textit{Group~C} & All & \textit{Group~A} & \textit{Group~B } & \textit{Group~C} \\ 
      \midrule
        Parents total & 115\,290 & 5\,386 & 106\,222 & 3\,682 & 115\,290 & 5\,386 & 106\,222 & 3\,682 \\ 
      \cmidrule{2-9} 
      Capellen & 9\,850 & 366 & 9\,244 & 240 & 9\,866 & 366 & 9\,256 & 244 \\ 
        Clervaux & 3\,885 & 190 & 3\,515 & 180 & 3\,872 & 190 & 3\,500 & 182 \\ 
        Diekirch & 6\,115 & 301 & 5\,516 & 298 & 6\,133 & 301 & 5\,528 & 304 \\ 
        Echternach & 3\,457 & 182 & 3\,161 & 114 & 3\,466 & 182 & 3\,176 & 108 \\ 
        Esch/Alzette & 34\,843 & 1\,721 & 31\,799 & 1\,323 & 34\,873 & 1\,721 & 31\,808 & 1\,344 \\ 
        Grevenmacher & 5\,799 & 246 & 5\,418 & 135 & 5\,804 & 246 & 5\,426 & 132 \\ 
        Lux. Campagne & 13\,411 & 486 & 12\,664 & 261 & 13\,424 & 486 & 12\,670 & 268 \\ 
        Lux. Ville & 18\,894 & 1\,132 & 17\,355 & 407 & 18\,850 & 1\,132 & 17\,336 & 382 \\ 
        Mersch & 6\,760 & 224 & 6\,323 & 213 & 6\,774 & 224 & 6\,320 & 230 \\ 
        Redange & 3\,752 & 132 & 3\,443 & 177 & 3\,754 & 132 & 3\,448 & 174 \\ 
        Remich & 3\,916 & 195 & 3\,585 & 136 & 3\,903 & 195 & 3\,582 & 126 \\ 
        Vianden & 981 & 47 & 893 & 41 & 973 & 47 & 890 & 36 \\ 
        Wiltz & 3\,627 & 164 & 3\,306 & 157 & 3\,598 & 164 & 3\,282 & 152 \\ 
      \midrule     
        Children total & 92\,515 & 6\,786 & 82\,489 & 3\,240 & 92\,518 & 6\,786 & 82\,488 & 3\,244 \\ 
          \cmidrule{2-9} 
        Capellen & 7\,933 & 481 & 7\,249 & 203 & 7\,929 & 475 & 7\,249 & 205 \\ 
        Clervaux & 3\,269 & 244 & 2\,842 & 183 & 3\,265 & 253 & 2\,835 & 177 \\ 
        Diekirch & 4\,823 & 366 & 4\,192 & 265 & 4\,923 & 379 & 4\,254 & 290 \\ 
        Echternach & 2\,782 & 232 & 2\,435 & 115 & 2\,813 & 226 & 2\,479 & 108 \\ 
        Esch/Alzette & 27\,769 & 2\,167 & 24\,435 & 1\,167 & 27\,751 & 2\,152 & 24\,424 & 1\,175 \\ 
        Grevenmacher & 4\,809 & 317 & 4\,371 & 121 & 4\,787 & 317 & 4\,360 & 110 \\ 
        Lux. Campagne & 10\,693 & 614 & 9\,845 & 234 & 10\,657 & 613 & 9\,824 & 220 \\ 
        Lux. Ville & 14\,773 & 1\,393 & 13\,046 & 334 & 14\,832 & 1\,398 & 13\,089 & 345 \\ 
        Mersch & 5\,444 & 284 & 4\,983 & 177 & 5\,393 & 280 & 4\,936 & 177 \\ 
        Redange & 3\,129 & 169 & 2\,808 & 152 & 3\,122 & 168 & 2\,798 & 156 \\ 
        Remich & 3\,141 & 246 & 2\,785 & 110 & 3\,130 & 244 & 2\,777 & 109 \\ 
        Vianden & 805 & 55 & 712 & 38 & 799 & 58 & 704 & 37 \\ 
        Wiltz & 3\,145 & 218 & 2\,786 & 141 & 3\,117 & 223 & 2\,759 & 135 \\ 
 \hline
   \end{tabular}
   }
  \end{table}

\section{Weighting and Calibration}\label{sec:weight}

In the sample, there is exactly one child $i$ assigned to each parent $k$.
Weighting can be applied after estimating response probabilities. The procedure is as follows:
We construct the vector of totals in the population
$$
\mathbf{t}_u = \begin{pmatrix} \sum_{k\in U} \mathbf{x}_k \\ \sum_{i \in V}\mathbf{z}_i \end{pmatrix}.
$$

For each parent $k$ selected along with their selected child $i$, we construct the vector
$$
\mathbf{u}_k = \begin{pmatrix} \mathbf{x}_k \\[2mm] \dfrac{ \mathbf{z}_{i} }{q_i r_i} \end{pmatrix}
= \begin{pmatrix} \mathbf{x}_k \\[2mm] m_k \pi_k \mathbf{z}_{i}  \end{pmatrix}.
$$
By virtue of the balancing equations \eqref{eq1} and \eqref{eq3}, we have
\begin{equation}
\sum_{k\in U} \frac{\mathbf{u}_k}{\pi_k} I_k \approx \mathbf{t}_u.
\label{eqbalu}
\end{equation}

However, due to nonresponse, the balancing equations~\eqref{eqbalu} are no longer satisfied.
Suppose that only a subset $R$ of the selected parents responded to the survey. Since the response mechanism depends only on the selected parent, the nonresponse adjustment is identical for both the responding parent and the corresponding sampled child.

If $S=\{k\in U\mid I_k=1\}$, we first estimate the response probability
\[
\widehat{\psi}_k = \widehat{\Pr}(k \in R \mid S),
\]
for example by fitting a logistic regression model using the auxiliary variables $\mathbf{u}_k$ as covariates.

Next, following the calibration approach of \citet{dev:sar:92}, we compute the calibration weights $g_k$ such that
\[
\sum_{k \in R} \frac{g_k \mathbf{u}_k}{\pi_k \widehat{\psi}_k}
\approx \mathbf{t}_u.
\]
 The resulting calibration weights can then be applied to both the parent and child estimators, thereby compensating for unit nonresponse while preserving calibration on the auxiliary variables.

The extrapolation weights for the parents are then
$$
w_k = \frac{g_k }{\pi_k \widehat{\psi}_k}, k\in R.
$$
The extrapolation weights for the children are
$$
w_i = \frac{g_k m_k }{ \widehat{\psi}_k}, i\in R,
$$
where child $i$ was selected by parent $k$.

\section{Conclusion}\label{sec:conclusion}

A key feature of the proposed design is that balancing is achieved simultaneously on auxiliary variables defined for both parents and children, thereby ensuring accurate estimation for the two target populations.
In summary, this complex sampling design enables to collect a balanced sample across diverse family contexts, including blended families. Stratification by pairs of co-responsibility and the random selection of one child per selected parent help overcome the difficulties associated with multiple co-responsibility relationships. The inclusion of auxiliary variables in the sampling ensures that the results can be accurately extrapolated for both parents and children. Although managing odd cycles in co-responsibility relationships poses a theoretical challenge, this situation is expected to be very rare in practice. This sampling design thus provides a novel solution for studies of modern families, yielding results that can be extrapolated to populations of parents and children, while respecting the constraints of balance and precision of the estimators. 

\section{Acknowledgments}
Authors would like to thank our colleague Audrey Bousselin  for her valuable comments and suggestions on earlier versions of this paper.

\bibliographystyle{apalike}
\bibliography{bibyves}

\end{document}